# Partitions of 4d Transition Metal Nuclei and Related Correlations Using the Core–Cluster Model

K. E. Abd El Mageed[1], A. G. Shalaby[2]

[1,2]Department of Physics-Faculty of Science- Benha university-Egypt

1- kmageed@yahoo.com

2- asmaa.shalaby@fsc.bu.edu.eg

**Abstract**

In the present work we attempt to study the cluster model in the transition metal region. The spectrum fitting method is studied for the selected nuclei ( $^{88,90,92}$Sr, $^{92,94}$Zr, $^{98,100}$Mo, $^{100,102,104,106}$Ru, $^{108,110}$Pd and $^{112,114,116,118}$Cd) with proton number ($38 \leq Z_T \leq 48$) and mass number ($88 \leq A_T \leq 118$). The core-cluster charge products are correlated with the transition probability B(E2↓; $2^+ \to 0^+$), the excitation energies and the product of valence nucleon numbers of the parent nuclei.

**I. Introduction:**

The idea of clustering (especially for $\alpha$-clustering) in light nuclei is established [1]. Recent development of the cluster model has been applied for near-magic cores and for some of heavier clusters which give a good results of the spectrum, electromagnetic properties and decay half-lives of many of heavy even-even nuclei in the rare earth and actinide elements [2, 3]. Depending on the cluster model, the cluster charge $Z_2$ correlates with the transition probability B (E2↓; 2+ → 0+), the excitation energies and the product of valence nucleon numbers of the parent nuclei [4]. Also The electric Quadrupole transition probability, B(E2)↑ for the first excited states, $2^+$ for even-even nuclei are listed [5]. The values of the transition probability are utilized to formulate the various systematic, empirical, and theoretical relationships that have been proposed in Ref. [5]. Recently, the electromagnetic transitions B(E2), B(M1) as a probe of clustering in nuclei [6, 7].

The first step to study the nuclei using the cluster model is to choose the appropriate core-cluster decomposition. For $^{20}$Ne, $^{44}$Ti, $^{94}$Mo and $^{212}$Po nuclei, many properties of the low – lying bands can be understood in terms of alpha clustering i.e. using core – cluster compositions in which both core and cluster are magic [8]. For example $^{88}_{38}\text{Sr} \to ^{84}_{36}\text{Kr} + ^{4}_{2}\text{He}$ then increasing the cluster charge ascending The exotic decay phenomenon in the actinide region leads to the choice of magic or near-magic residual nuclei as cores, and the emitted nuclei as clusters [9].

Buck et al., [4, 10] focused on the study of the ground state bands of even-even nuclei in the rare - earth and actinide regions. In Ref. [11] the spectra of even -even nuclei and the transition probability have been studied for tetravalence transition element isotopes Hf. In the present work, we have chosen and studied the ground -state band of some even-even nuclei and some of their isotopes ($^{88,90,92}$Sr, $^{92,94}$Zr, $^{98,100}$Mo, $^{100,102,104,106}$Ru, $^{108,110}$Pd and $^{112,114,116,118}$Cd) in the transition metal region using the cluster model.

The outline of the paper is as follows, In section II we discuss the cluster model beginning with the selection of the core cluster as even -even and the study of their spectra. Section III presents the core cluster charge and the correlation relation to different quantities of interest. And section IV contains the results and discussion.

**II. The used Model**

**II.1 Choice of the Core – Cluster**

The general method of selecting core and cluster is, to make the possible which should give preference to combinations that have internal stability [12]. So that, an even – even nucleus of mass and charge ($A_T$, $Z_T$) should be split into an even-even, core ($A_1$, $Z_1$) and cluster ($A_2$, $Z_2$).. The function D (1, 2) is maximized as,

$$D(1, 2) = \left[B_A(Z_1, A_1) - B_L(Z_1, A_1)\right] + \left[B_A(Z_2, A_2) - B_L(Z_2, A_2)\right] \quad (1)$$

Where, (1, 2) refer to the core and the cluster respectively, and $B_A$ is an actual binding energy and $B_L$ is a liquid drop binding energy given by,

$$B_L = a_v A_T - a_s A_T^{2/3} - a_c \frac{Z_T^2}{A_T^{1/3}} - a_a \frac{(A_T - 2Z_T)^2}{A_T} + \delta, \quad (2)$$

All the parameters [13] in eq. 2 are listed in table( 1).

**Table (1) Parameters in eq. 2**

| Parameter | Value (MeV) |
|---|---|
| $a_v$ | 15.56 |
| $a_s$ | 17.23 |
| $a_c$ | 0.697 |
| $a_a$ | 23.285 |
| $\delta$ | $12/A_T^{1/2}$ |

For the selected nucleus, when the conditions $A_1 = A_T - A_2$ and $Z_1 = Z_T - Z_2$ are applied, D remains a function of two independent variables, the cluster mass and charge, ($A_2$, $Z_2$). A simple formula of D(1, 2) results from observation that the electric dipole transitions between low-laying bands of opposite parity in heavy nuclei are very weak. This implies that, the total nuclear mass and charge should be distributed in the same proportions between core and cluster resulting in the no-dipole constraint [12],

$$\frac{<Z_1>}{A_1} = \frac{<Z_2>}{A_2} = \frac{Z_T}{A_T} \qquad (3)$$

In eq. (3) the values of ($A_1$, $Z_1$) and ($A_2$, $Z_2$) are non-integral, which can be interpreted as average of charges arising from a suitable weighted mixtures of cores and clusters [2, 12].

**II.2 Spectra of Even-Even Nuclei**

The spectra of the even–even nuclei are studied using the Bohr-Sommerfeld relation [14]

$$\int_{r_1}^{r_2} dr \sqrt{\frac{2\mu}{\hbar^2}\left[E(G,L) - \left[V_N(r,R) + V_C(r,R) + \frac{\hbar^2(L+0.5)^2}{2\mu r^2}\right]\right]} = (G-L+1)\frac{\pi}{2} \qquad (4)$$

Where $r_1$ and $r_2$ are the two inner turning points, μ is the relative mass of the core-cluster decomposition and G is the global quantum number identifying a band levels given by

$$G = 2n + L \qquad (5)$$

Where L is the angular momentum of a particular level and, n is the number of nodes in the radial wave function of the cluster-core relative motion. For the transition metals region we set $G = 4A_2$, with $A_2$ the cluster mass, is appropriate in the present calculations. The core-cluster interaction in equation (4) includes nuclear and coulomb terms. For the nuclear interaction $V_N(r, R)$, we use the modified Woods-Saxon potential with standard parameter values given by [15],

$$V_N(r,R) = -\left(\frac{A_1 A_2}{A_T}\right) V_0 \frac{f(r,R,x,a)}{f(0,R,x,a)}$$

$$f(r,R,x,a) = \left[\frac{x}{(1+\exp[(r-R)/a])} + \frac{1-x}{(1+\exp[(r-R)/3a])^3}\right] \qquad (6)$$

And $V_0$ = 57 MeV, $x$=0.33, and $a$ =0.74 fm. And the Coulomb potential $V_C(r, R)$ works between a uniformly charged spherical core of radius R, and a point cluster, then the core cluster interaction can be completely defined. The value of R is only free parameter in the calculation.

## III  Core-Cluster Charge and Related Correlation

In this section we discuss the relation between the core-cluster charge product ($Z_1Z_2/Z_T$), and different related correlation like the transition probability, B (E2; $2^+ \rightarrow 0^+$) [16]. It is given by,

$$B(E2) = \frac{1}{4\pi}\left(\left(\frac{Z_1Z_2}{Z_T}\right)\int \chi_2(r) \, r^2 \chi_0 dr\right)^2 \tag{7}$$

Where: $\chi_\lambda(r)$ is the radial wave function of the core-cluster relative motion for angular momentum $\lambda$. The radial Schrödinger equation for $\chi_\lambda(r)$ is written as,

$$\frac{-\hbar^2}{2\mu}\frac{d^2\chi_\lambda}{dr^2} + V(r)\chi_\lambda + \frac{\hbar^2 \lambda(\lambda+1)}{2\mu R^2}\chi_\lambda = E_\lambda \chi_\lambda \tag{8}$$

The solution of the Schrödinger equation (8) one can get the energy states and the wave function.. near-identical $\chi_\lambda(r)$ [17], the Solution of eq. (7) can be written as,

$$\frac{\sqrt{B(E2)}}{A_T^{2/3}} = \frac{1}{\sqrt{4\pi}} r_0^2 \left\{\frac{Z_1Z_2}{Z_T}\right\} \tag{9}$$

Where V (r) is the core-cluster potential and $\mu = A_1A_2/A_T$ is the reduced mass. The solution of equation (8) are discussed in details in ref. [4]

So that

$$\left(\frac{Z_T}{(E_L - E_\ell)A_T^{5/3}}\right) = \frac{r_{L\ell}^2}{d_{L\ell}}\left(\frac{Z_1Z_2}{Z_T}\right) \tag{10}$$

Where:

$$d_{L\ell} = \frac{\hbar^2}{2}\left[L(L+1) - \ell(\ell+1)\right] \tag{11}$$

In the present work we take the transition between the levels that have, ($L = 2$, $\ell = 0$) and, ($L = 4$, $\ell = 2$).

Buck and et al. [4] also relate the charge products ($Z_1Z_2/Z_T$) to the products $N_PN_N$. Here $N_P$ and $N_N$ are the number of valence protons and valence neutrons, respectively. The $N_pN_n$ scheme and related different correlations are discussed [18]. In the near-magic core, so that the cluster has $Z_2 \approx Z_p$ and $N_2 \approx N_N$, and the core has $Z_1 \approx Z_T$ and $N_1 \approx N_T$. For the no-dipole constrain [4] of eq.(3),

$$N_P N_N = Z_2 N_2 = Z_2^2 \frac{N_T}{Z_T} \approx \left(\frac{Z_1 Z_2}{Z_T}\right)^2 \frac{N_T}{Z_T}$$

So that

$$\sqrt{\frac{Z_T}{N_T}} (N_P N_N)^{1/2} = \left(\frac{Z_1 Z_2}{Z_T}\right) \qquad (12)$$

Using the standard shell closures, the values of $N_p$ and $N_N$ are calculated at nucleon number 50 and 82. By studying the selected nuclei with ($38 \leq Z_T \leq 48$) using eq. (12), and to overcome the difficulties which reproduce from the selection of certain proton subshell closures, one has to generalize the mentioned equations above as reported [4, 10] to be as following;

$$\frac{\sqrt{B(E_2)}}{A_T^{2/3}} = a_0 + \frac{1}{\sqrt{4\pi}} r_0^2 \left(\frac{Z_1 Z_2}{Z_T}\right) \qquad (13)$$

$$\left(\frac{Z_T}{(E_L - E_\ell) A_T^{5/3}}\right) = a_{L\ell} + \frac{r_{L\ell}^2}{d_{L\ell}} \left(\frac{Z_1 Z_2}{Z_T}\right) \qquad (14)$$

$$\sqrt{\frac{Z_T}{N_T}} (N_P N_N)^{1/2} = \alpha + \beta \left(\frac{Z_1 Z_2}{Z_T}\right) \qquad (15)$$

In the present work, we have applied Eqs. (13- 15) to the present data for $A_T \leq 118$, using the averaged $Z_2$ values from table (2).

## IV. Results and Discussion

In the present work we selected the appropriate core- cluster decomposition for all the chosen nuclei ($^{88-92}$Sr, $^{92, 94}$Zr, $^{98, 100}$Mo, $^{102-106}$Ru $^{108, 110}$Pd, $^{112-118}$Cd) in the transition metals region in the periodic table by using equations (1&3). The binding energies for different nuclei and isotopes as well are tabulated [19]. . We calculated D (1, 2) as a function of the cluster charge $Z_2$ from equation (1). Figure(1 ; a, b, c, d, e, and f) presents the plotting of the function D(1, 2) versus the cluster charge $Z_2$ for different transition elements and their isotopes. It is clear that from fig.(1), the suitable cluster is Beryllium (Be) nucleus in which $Z_2 = 4$ for the chosen nuclei. Therefore, we calculated the average of the cluster charge $<Z_2>$ based on the calculations of eqs. (1 & 3) as listed in table (2).

**Table (2):-** The cluster charge $Z_2$ calculated by different methods discussed in the text and their average values.

| Nucleus | $Z_2$ from eq. (3) | $Z_2$ from D(1,2) | Average $<Z_2>$ |
|---|---|---|---|
| $^{88}$Sr | 2.59 | 4 | 3.295 |
| $^{90}$Sr | 2.53 | 4 | 3.265 |
| $^{92}$Sr | 2.478 | 4 | 3.23913 |
| $^{92}$Zr | 2.6087 | 4 | 3.30435 |
| $^{94}$Zr | 2.5532 | 4 | 3.2766 |
| $^{98}$Mo | 3.428 | 4 | 3.714 |
| $^{100}$Mo | 4.2 | 4 | 4.1 |
| $^{100}$Ru | 4.4 | 4 | 4.2 |
| $^{102}$Ru | 4.314 | 4 | 4.157 |
| $^{104}$Ru | 5.07 | 4 | 4.535 |
| $^{106}$Ru | 4.9 | 4 | 4.45 |
| $^{108}$Pd | 5.111 | 4 | 4.5555 |
| $^{110}$Pd | 5.02 | 4 | 4.51 |
| $^{112}$Cd | 5.14 | 4 | 4.57 |
| $^{114}$Cd | 5.05 | 4 | 4.525 |
| $^{116}$Cd | 5.79 | 4 | 4.895 |
| $^{118}$Cd | 5.695 | 4 | 4.8475 |

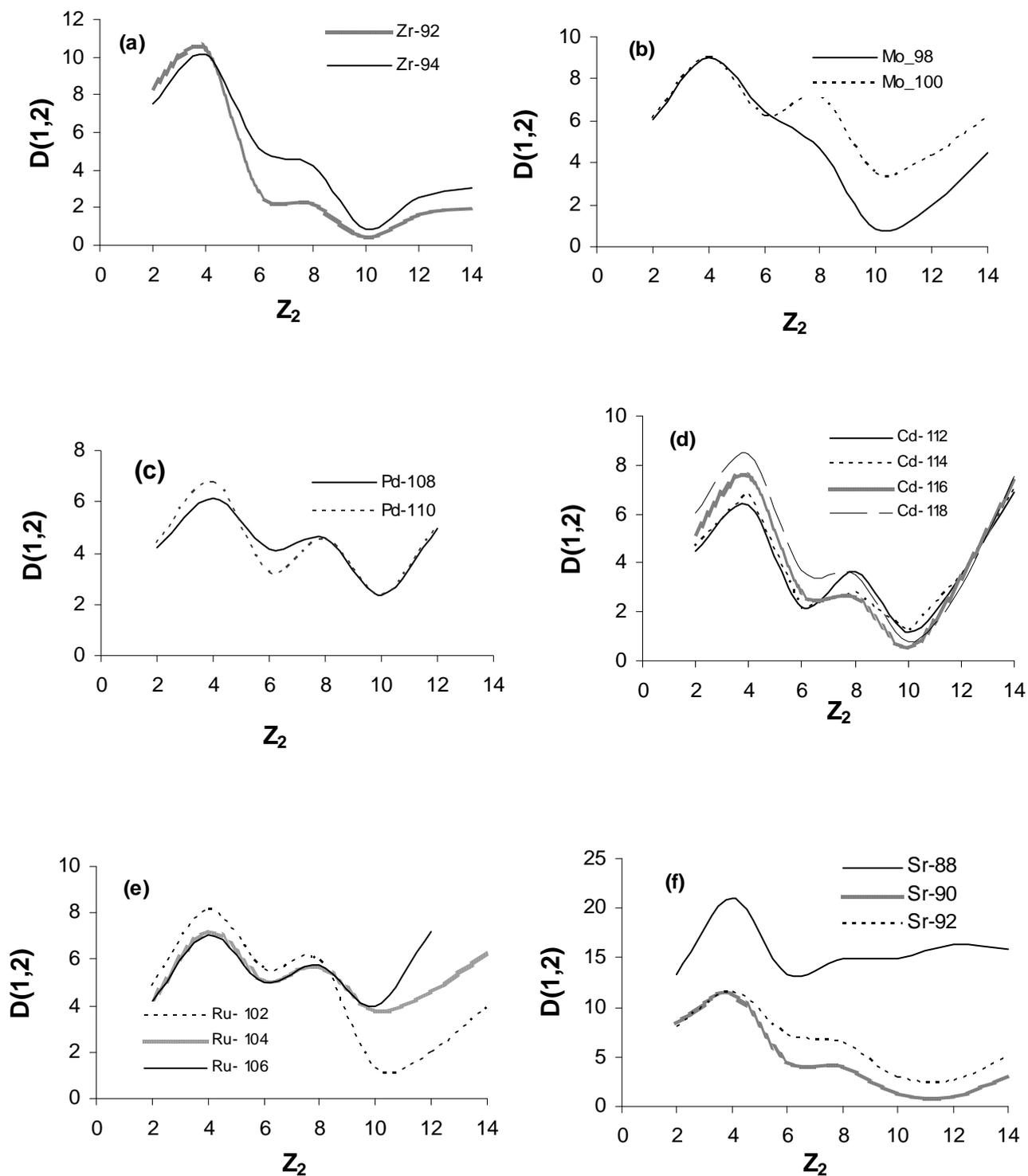

**Fig. (1):- Calculations of D(1,2) as a function of cluster charge $Z_2$ for different transition metals nuclei and their isotopes. (a) Zr $^{92, 94}$, (b) Mo$^{98, 100}$, (c) Pd $^{108, 110}$, (d) Cd $^{112, 114, 116, 118}$, (e) Ru $^{102, 104, 106}$, (f) Sr $^{88, 90, 92}$**

We have used equation (4) to calculate the energy levels of the ground state band for the chosen nuclei. Figures (2-6) show the comparison between the calculated energy levels and the available experimental data [20] versus spin. Using the considered partition, the calculated energy levels give a satisfied agreement with the experimental data for the ground-state band.

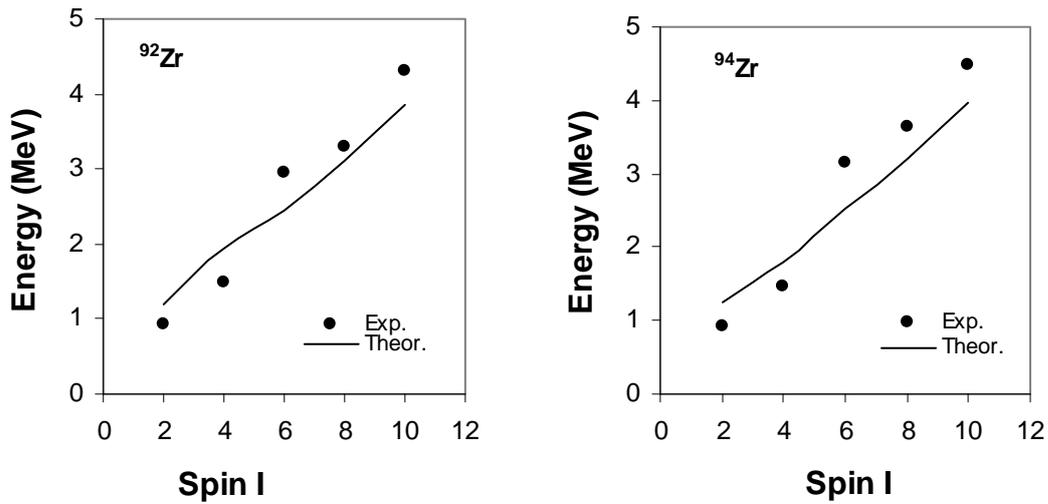

**Fig. (2):- The energy spectra versus the spin for even nuclei $^{92,94}$Zr, compared with the experimental data [20].**

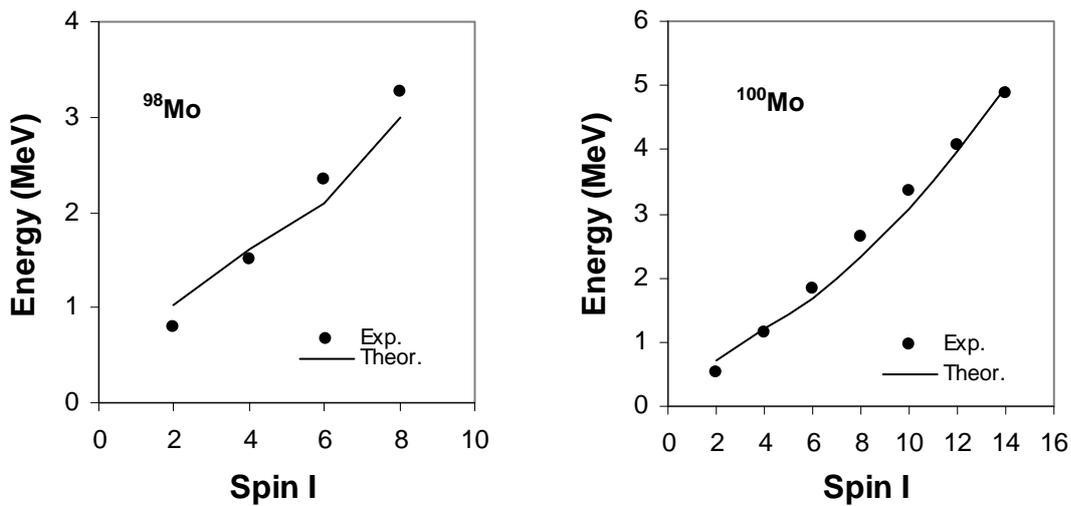

**Fig. (3):- The energy spectra versus the spin for even nuclei, $^{98,100}$Mo, compared with the experimental data [20].**

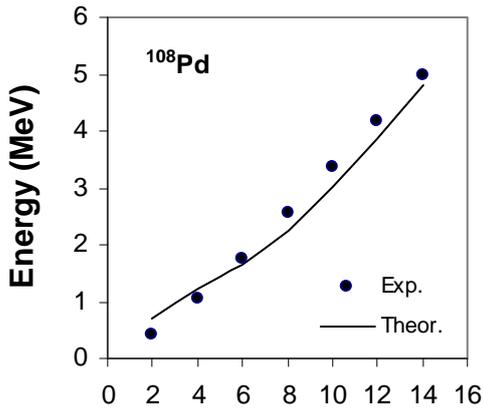
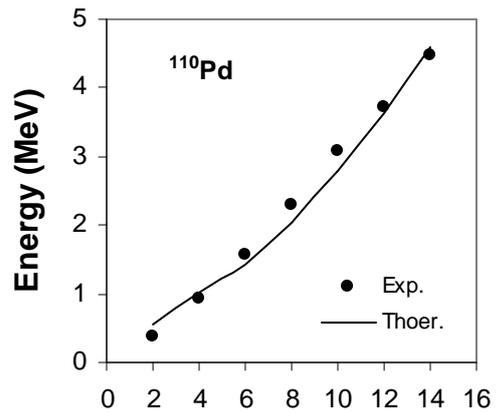

**Fig. (4):-** The energy spectra versus the spin for even nuclei, $^{108,110}$Pd compared with the experimental data [20].

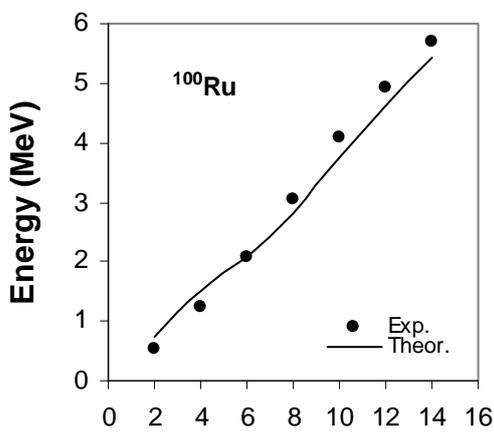
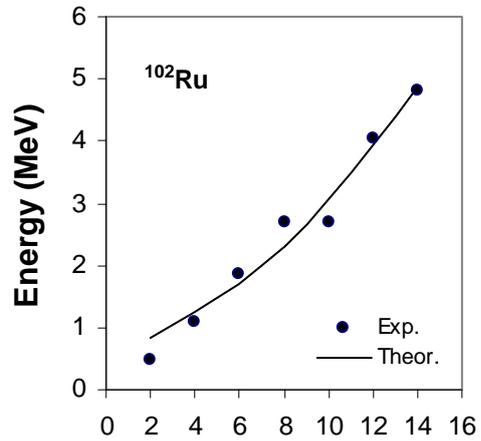

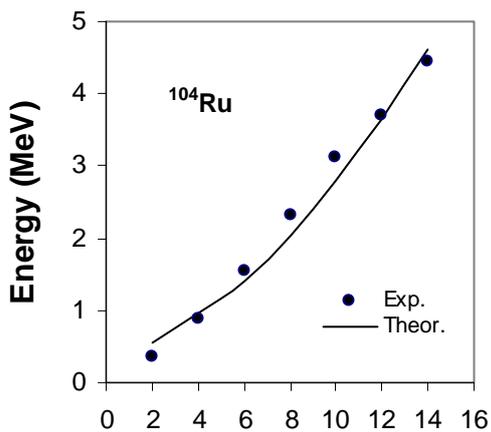
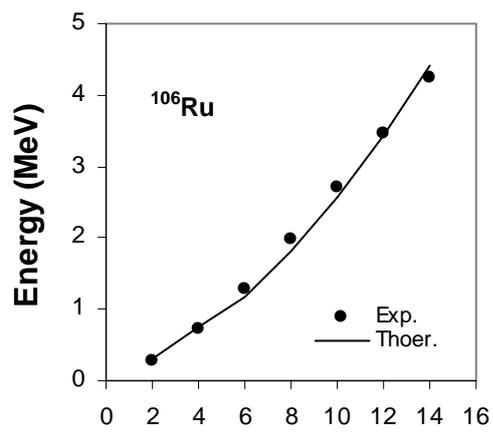

**Fig. (5):-** The energy spectra versus the spin for even nuclei $^{100,102,104,106}$Ru, compared with the excited states experimental data [20].

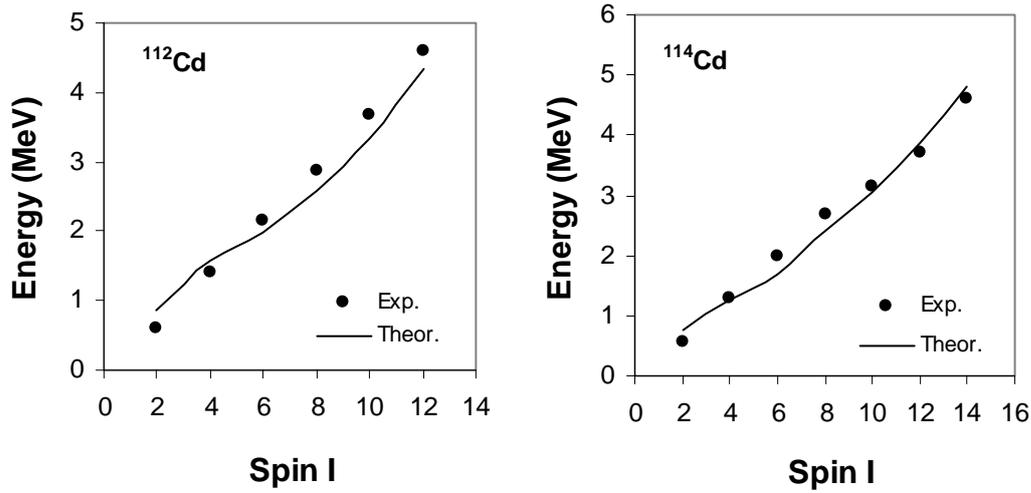

**Fig. (6):- The energy spectra versus the spin for even nuclei $^{112,114}$Cd, compared with the experimental data [20].**

Figure (7) shows the plot of measured values [19] of [B (E2) in $e^2$ fm$^2$] with error bars versus the core-cluster charge product $(Z_1Z_2/Z_T)$ according equation (13). The solid line is the best linear fit of the data, the gradient equals $1.6414 \pm 0.24976$ of the data and the intercept equals $a_0$ = $-3.21269 \pm 0.93231$. So, $a_0$ in equation (13) has negative value for the transition nuclei. And the value of $r_0$ can be calculated. The gradient equals $\frac{1}{\sqrt{4\pi}} r_0^2$, so that $r_0 = 2.4121 \pm 0.9409$ fm.

Figure (8) Shows the plot of the of the correlation relation eq. (14) versus $(Z_1Z_2/Z_T)$ with L=2,l=0. From this figure, the gradient = $0.01646 \pm 0.00519$ and the intercept equal $-0.02598 \pm 0.01938$. From these values we can determine the following values $r_{20} = 0.222 \pm 0.124$ fm, $a_{20} = 0.02598 \pm 0.01938$.

Figure (9) shows the same plot as eq. (14) versus the ratio $(Z_1Z_2/Z_T)$ with L= 4, l=2. From this figure, the gradient equal $0.0017804 \pm 0.00375$ and the intercept equal $0.02246 \pm 0.01403$. From these values we can determine the following values, $r_{42} = 0.11 \pm 0.16$ fm, and $a_{42} = 0.022466 \pm 0.01403$. In ref. [4], they removed some points from the same figures according to the cluster charge $Z_2 \leq 6$ so that $r_{20}, r_{42}$ are consistent with each other.

In the present work, it is clear that the values of $r_{20}, r_{42}$ are consistent with each other and there is no need to remove any points from figures (8, 9). Figure (10) Shows the results of plotting eq. (15) with the fitted line corresponding to $\alpha = 0.8956 \pm 0.8494$ and $\beta = 2.084 \pm 3.171$.

Substituting $[N_P N_N Z_T/N_T]^{1/2}$ for $(Z_1 Z_2 /Z_T)$ in equations (13) and (14) generates linear plots similar to those of Figures (7), (8) and (9).

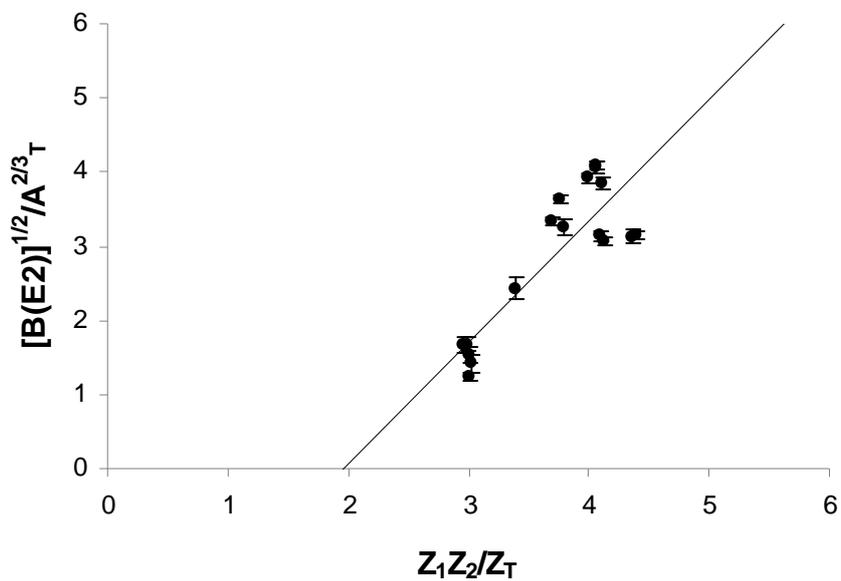

**Fig. (7) The experimental values of $[B(E_2)]^{1/2}$ $A_T^{2/3}$ with error bars versus the ratio $Z_1Z_2/Z_T$**

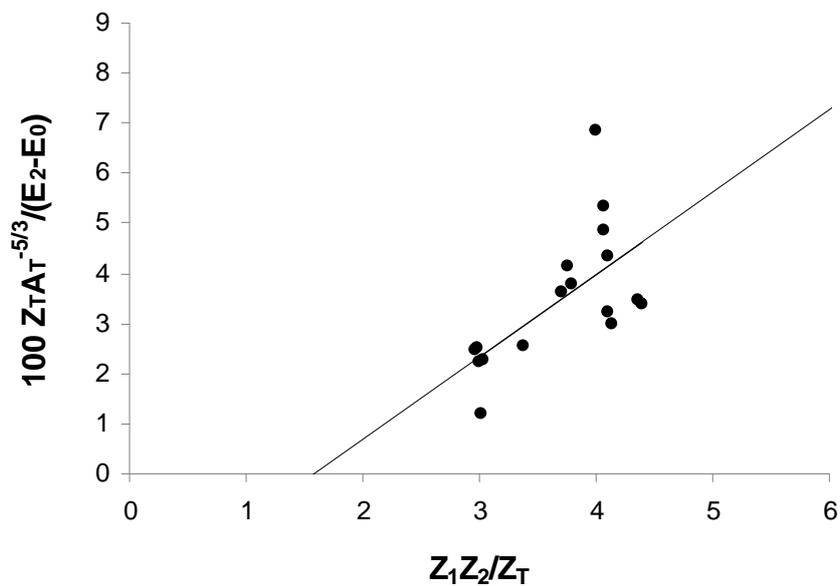

**Fig (8) The values of $100\, Z_T A_T^{-5/3}/(E_2 - E_0)$ versus the ratio $Z_1Z_2/Z_T$, the solid line is the best linear fit**

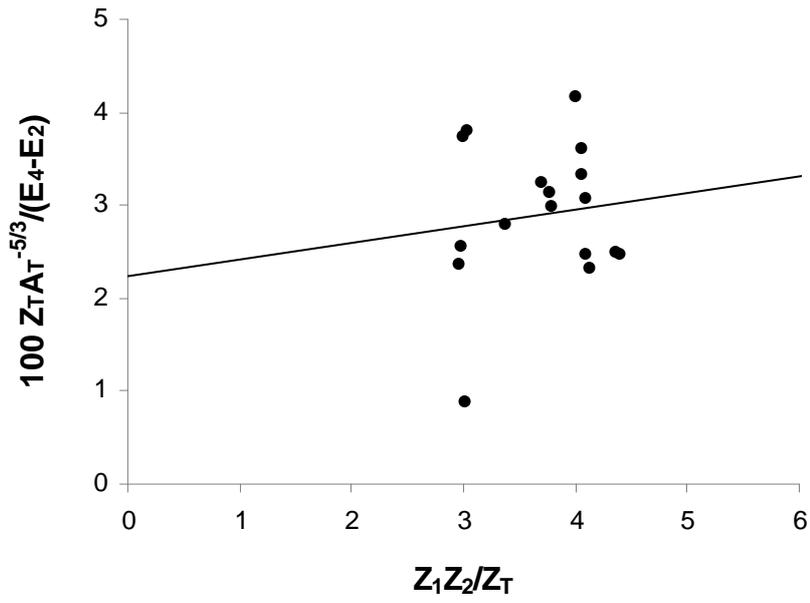

**Fig (9)** The values of $100\, Z_T A_T^{-5/3}/(E_4 - E_2)$ versus the ratio $Z_1 Z_2/Z_T$, the solid line is the best linear fit

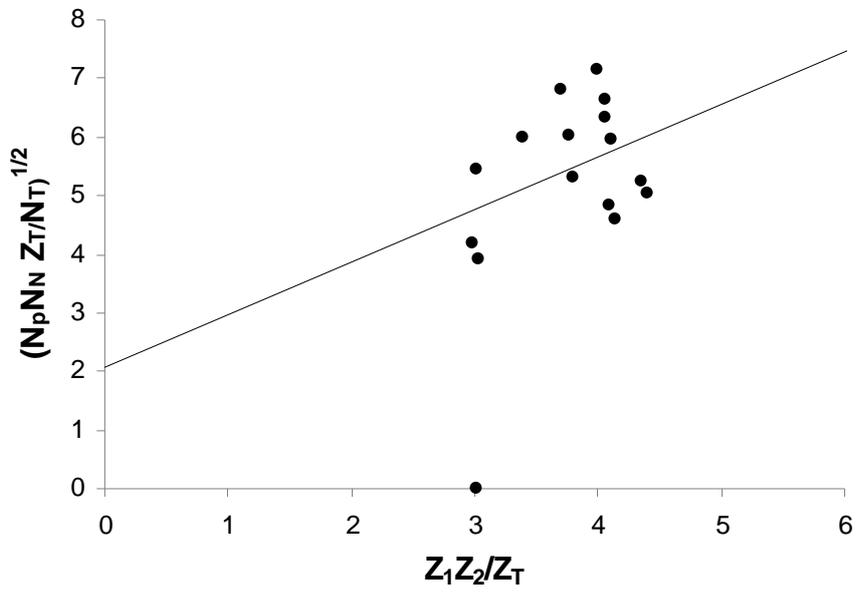

**Fig. (10)** Calculations of $\left( N_p N_N Z_T / N_T \right)^{1/2}$ versus the ratio $Z_1 Z_2/Z_T$, the solid line is the best linear fit

## V. Conclusion

The applicability of using the core-cluster model in the light , intermediate and some heavy nuclei have succeed. In the present work, we have investigated also the applicability of using the core - cluster model in the transition metal elements from the periodic table. We have got a good results in describing the different correlation between the core cluster charge product, and the transition probability B (E2↓; $2^+ \rightarrow 0^+$), the excitation energy and the number of valence protons and neutrons for different states. From this work we can conclude the following:

1 - The core- cluster (even-even) compositions from the calculation the function D(1, 2) and eq. (3) were reported. We have got the suitable cluster charge is 4 for Beryllium (Be). Then the average of the cluster charge calculated.

2- Studying the energy levels of the ground - state band for the chosen nuclei has shown a satisfied agreement with the experimental data.

3- The intercepts in the correlation relation eq. (14) are different signs. In the case of the transition $2^+ \rightarrow 0^+$ is negative while in case of $4^+ \rightarrow 2^+$ is positive. We suggest this difference due to the values of the excitation energy states.